\documentclass[12pt]{iopart}
\usepackage{graphicx}
\usepackage{iopams}
\expandafter\let\csname equation*\endcsname\relax
\expandafter\let\csname endequation*\endcsname\relax
\usepackage{amsmath}
\usepackage{url}
\usepackage{bbold}
\usepackage{cite}

\usepackage[caption=false]{subfig}
\usepackage{overpic}

\begin{document}
\title{A didactic approach to quantum machine learning with a single qubit}

\author{Elena Pe\~{n}a Tapia$^1$, Giannicola Scarpa$^2$, and Alejandro Pozas-Kerstjens$^3$}

\address{$^1$ Universidad Polit\'ecnica de Madrid, Madrid, 28031, Spain}
\address{$^2$ Escuela T\'enica Superior de Ingenier\'ia de Sistemas Inform\'aticos,
Universidad Polit\'ecnica de Madrid, Madrid, 28031, Spain}
\address{$^3$ Instituto de Ciencias Matem\'aticas (CSIC-UAM-UC3M-UCM), 28049 Madrid, Spain}

\begin{abstract}
This paper presents, via an explicit example with a real-world dataset, a hands-on introduction to the field of quantum machine learning (QML).
We focus on the case of learning with a single qubit, using data re-uploading techniques.
After a discussion of the relevant background in quantum computing and machine learning we provide a thorough explanation of the data re-uploading models that we consider, and implement the different proposed formulations in toy and real-world datasets using the \texttt{qiskit} quantum computing SDK.
We find that, as in the case of classical neural networks, the number of layers is a determining factor in the final accuracy of the models. Moreover, and interestingly, the results show that single-qubit classifiers can achieve a performance that is on-par with classical counterparts under the same set of training conditions. While this cannot be understood as a proof of the advantage of quantum machine learning, it points to a promising research direction, and raises a series of questions that we outline.

\end{abstract}

\maketitle

\section{Introduction}
Quantum Machine Learning (QML) has become one of the hottest trends in the area of quantum computing and quantum information science of 2022.
However, QML is also a field in the making, so much so that it is difficult to find a consensual ``official'' definition of the term. In its most general sense, quantum machine learning explores synergies between \textbf{quantum computing} and \textbf{machine learning} \cite{biamonte2017quantum,schuld2018supervised}.
This broad definition for QML can apply to a wide variety of approaches, from the use of machine learning in experimental physics \cite{denby1990,baldi2014,abiuso2022,chen2022} to quantum-inspired machine learning \cite{stoudenmire, arrazola2020,felser2021,pozas2022}, including the approach we focus on for this paper: the development of quantum algorithms for machine learning tasks, with the intention of potentially \textit{``learning better''}.
This leads to another point to consider, and that is \textit{what does ``learning better'' mean}, and the answer is not easy either. The concept of ``quantum advantage'' (how can a quantum computer outperform its classical counterpart) is still an open question itself, and while some works have proven certain advantages in specific learning problems, these are still quite narrow, and heavy considerations are taken into account.

Despite these pending questions, there are reasons to believe that combining quantum computing and machine learning can still be a very interesting path to pursue, both from the theoretical and practical point of view. As mentioned in Ref.~\cite{arunachalam2017guest}, machine learning models are not theoretically well understood, their advantage is not necessarily mathematically proven, and still they are able to perform amazingly well in a wide variety of tasks. At the same time, the authors of Ref.~\cite{schuld2018supervised} point out how the field of machine learning is suffering from a lack of fundamentally new research directions. By studying its intersection with quantum computing, new insights could arise to improve our current understanding of these models and open new ways for improving their performance and efficiency.

The motivation behind this paper arises from the study of QML models from the point of view of a potential ``customer'', trying to look for the most appropriate model for a specific commercial application.
Most current efforts in QML are centered in algorithms that can be implemented in currently available quantum devices and still provide some relevant learning capability.
In this context of NISQ-focused implementations \cite{preskill2018quantum}, the proposal of P\'{e}rez-Salinas et al. \cite{perez2020data,perez2021one} stands out for its simplicity: the authors claim that single-qubit circuits can be used as universal quantum classifiers and learn effectively on multi-dimensional data. From a practical perspective, minimizing the quantum resources involved (qubits and quantum operations) seems like a logical direction to follow. However, when facing a real-world task, how well can a single-qubit neural network perform?

In order to answer this question, and to provide further insight into the advantages and practical challenges of single-qubit learning, this paper focuses on implementing and testing the premise in a real-world task, in particular, fraud detection using credit card transaction data.
For these means, we analyze different single-qubit architectures proposed in Refs. \cite{perez2020data} and \cite{perez2021one} and present a novel implementation using the \texttt{qiskit} SDK \cite{Qiskit}. This \texttt{qiskit} code is later used in two experimental setups: we first test these architectures on a toy dataset to verify the theoretical results from Ref.~\cite{perez2020data}. We then proceed to apply the insights obtained from the first experiments to train a single-qubit QNN model on real data, and benchmark it against a classical neural network algorithm. Because of its straightforward application and availability, we decide to perform this second set of experiments on the Kaggle credit card fraud detection dataset \cite{kaggle}.

It is worth noticing that recent independent work demonstrates the growing interest of the scientific community into both the practical availability single-qubit quantum neural networks and the industrial applications of QML.
For instance, Ref.~\cite{perez2022ion} implements single-qubit classification using ion-trap quantum devices, while Ref.~\cite{kyri2022frauddet} uses more complex QNN models on the Kaggle fraud detection dataset.

\subsection{Organisation of the paper}
The remainder of this paper is structured as follows: in Section \ref{chap:preliminaries} we introduce the necessary technical concepts to understand the theory behind the implementation of single-qubit neural network algorithms; in Section \ref{chap:algorithm} we dive deeper into the analysis of the single qubit as a universal approximator; finally, in Sections \ref{chap:experiment} and \ref{chap:conclusion}, we present the main experimental setup and results for the assessment of this QNN architecture in a real-world scenario, respectively, as well as a brief final discussion.

\section{Technical Preliminaries}
\label{chap:preliminaries}
In this section we introduce the basic theoretical concepts necessary to understand the logic behind the single-qubit quantum classifier and other algorithms based on quantum neural networks. These concepts belong to the two parent disciplines of quantum machine learning: machine learning and quantum computing. For an in-depth study of these two fields we recommend books \cite{nielsen2002quantum} and \cite{nielsen2015neural}.

\subsection{Machine learning}
Machine learning is a branch of Artificial Intelligence (AI) that covers algorithms able to learn from data, in an attempt to imitate the way humans learn and generalize from experience \cite{massoli2021leap,ibmblogml}. Artificial Neural Networks and Kernel Methods are the main algorithmic families within Machine Learning.

Neural Networks (NNs) are a biologically-inspired programming paradigm that enables computers to learn from observational data \cite{nielsen2015neural}. The building blocks of neural networks are a series of computational nodes known as \textit{(artificial) neurons}.
There are several artificial neuron models, but one of the most widely used is the \textit{sigmoid} or logistic neuron. This type of neuron takes a series of normalized inputs $x_j$ (with values between 0 and 1) and computes a parametrized sum $\sum_jw_jx_j + b$, where $w_j$ are known as weights and $b$ is the bias. This parametrized sum is then inputted into the neuron's \textit{activation function}:
\begin{equation*}
    a = \sigma\left(\sum_jw_jx_j + b\right),
\end{equation*}
where $\sigma(z)$ is typically the \textit{sigmoid function}, defined as $\sigma (z) = \left(1+e^{-z}\right)^{-1}$.
Intuitively, this formulation can be seen as a way for the neuron to  ``weigh evidence (inputs) to make decisions (output)'', as cleverly described in Ref.~\cite{nielsen2015neural}. The neuron can \textit{learn} by optimizing the weights $w_j$ and bias $b$ to produce a specific output (decision) for a specific set of inputs (evidence). The role of the activation function is to ``moderate'' this learning process, so that small changes in the learnable parameters (weights and bias)  cause small changes in the output.

The optimization of the weights and biases is typically performed via gradient descent: for a specific task, a measure of the error between the outputs produced by the neural network and the expected output (often known as \textit{loss} or \textit{cost} function) is computed, and the weights and biases are updated in the direction of the corresponding gradients of this error. The computation of the gradients is especially easy for the sigmoid function, since it satisfies

\begin{equation}
    \frac{\partial \sigma(z)}{\partial z} = \sigma(z)\left[1-\sigma(z)\right].
\end{equation}

This intuition can also be applied to other activation functions besides the sigmoid, with changes in the way the partial derivatives are calculated.

Neural network architectures are comprised of a series of interconnected \textit{layers}, each of which can contain one or more neurons. An example can be found later in the article, in Figure \ref{fig:nn}. For example, consider a shallow neural network, with three distinct layers: an input layer, used for encoding the inputs; an intermediate hidden layer; and an output layer with a single neuron, used to process the network's output. Networks where the output from one layer is used as input for the following one are called \textit{feed-forward networks}. The most common NN architectures follow this model, but it should be noted that some types of neural networks, such as \textit{recurrent neural networks}, do not, as they include feedback loops.

\label{sec:uat}
The reason why the neural network models are so powerful and widely used is that even the most basic architecture (one hidden layer) is able to approximate any continuous function up to a certain precision. If the number of hidden layers is increased, the precision of the approximation will increase too. This is known as the \textit{Universal Approximation Theorem (UAT)}.
There is a series of mathematical proofs of this theorem \cite{cybenko1989approximation,hornik1989multilayer} behind the scope of this brief conceptual introduction. Nielsen provides a very useful visual explanation of this theorem in Chapter 4 of his online book \cite{nielsen2015neural}.

The idea behind neural network training can also be seen from a different perspective. A specific neural network architecture can be understood as a continuous parametrization of a family of models. Then, the process of training consists of choosing, among the family, the model that better achieves the target task. In this spirit, training a machine learning model seems not to differ from regression. The crucial difference between both is that, while in regression the target task is finding the function that better fits the available data, in machine learning the task is finding the function that better performs \textit{in datapoints different than those available for training}.

The learning paradigm we use in this paper is supervised learning. It uses a set of pairs of input and corresponding desired output, with the goal of learning how to produce the desired output for each input. The cost function in supervised learning is related to eliminating incorrect predictions, or reducing the error. We will concentrate on classification models, where the output data is a discrete class label indicating the category to which the input belongs.

\subsection{Quantum computing}
Quantum computing is a sub-field of quantum information processing that leverages quantum mechanical phenomena to enable complex computations effectively. This section will introduce the core computational concepts necessary to understand the mechanisms behind the single-qubit quantum neural network.

The \textit{qubit} is the fundamental unit of information of quantum device, a quantum variant of the classical bit. From the quantum-mechanical perspective, a qubit is a quantum system associated with two measurable events, commonly represented as the Dirac vectors $|0\rangle$ and $|1\rangle$.
These two vectors form an orthonormal basis of a 2-dimensional Hilbert space (a complex vector space with an inner product).
This is known as the \textit{computational basis}, and $|0\rangle$ and $|1\rangle$ will be referred to as the \textit{computational basis states}.

The most general way of representing a qubit's state $|\psi\rangle$ is as a \textit{superposition}: this is a normalized linear combination of the basis states with complex coefficients. These superpositions are mathematically defined as \textit{state vectors} of the form:
\begin{equation}
    |\psi\rangle = \alpha_0 |0\rangle + \alpha_1 |1\rangle \quad\text{such that}\quad  \alpha_0, \alpha_1 \in \mathbb{C},\,
    |\alpha_0|^2 + |\alpha_1|^2 = 1.
    \label{eq:statevec}
\end{equation}

Equivalently, in vector notation we have:
\begin{gather*}
|0\rangle = \begin{pmatrix}1\\0\end{pmatrix} \in \mathbb{C}^2; \;\;\;
|1\rangle = \begin{pmatrix}0\\1\end{pmatrix} \in \mathbb{C}^2\\
|\psi\rangle = \begin{pmatrix}\alpha_0\\\alpha_1\end{pmatrix}
\end{gather*}

Superposition is one of the key quantum properties of the qubit, since it allows for it to simultaneously exist in states $|0\rangle$ and $|1\rangle$ in absence of measurement. When measured, the state of the qubit \textit{collapses} to one of the two basis states with different probabilities, which correspond to the square of the complex coefficients $\alpha_0$ and  $\alpha_1$, known as \textit{amplitudes}.

Equation \ref{eq:statevec} can be rewritten in the following, more convenient form for representing a generic single qubit's state:
\begin{gather*}
    |\psi\rangle = e^{\mathrm{i}\gamma}\left[ \cos\left(\frac{\theta}{2}\right) |0\rangle + e^{\mathrm{i}\phi} \sin\left(\frac{\theta}{2} \right) |1\rangle \right] \\
    0\leq\theta\leq\pi \in \mathbb{R} \\
    0\leq\phi,\gamma\leq 2\pi \in \mathbb{R},
\end{gather*}
where the parameter $\gamma$ represents the global phase of the qubit, and can be omitted for the sake of this analysis, since it does not have any measurable effect on the single qubit's state. The remaining parameters, $\theta$ and $\phi$, can be seen as spherical coordinates that allow to map the state of the qubit into a point in the surface of a sphere of radius 1. This sphere is known as the \textit{Bloch sphere}, and it's essentially a geometrical representation of the qubit's state, whose coordinates are characterized by the following \textit{Bloch vector}:
\begin{gather*}
    |\psi\rangle = (\sin\theta \cos\phi, \sin\theta \sin\phi, \cos\theta) \in \mathbb{R}^3
\end{gather*}

In future sections we will cover how single-qubit operations can be seen as rotations within the Bloch sphere with two degrees of freedom, corresponding to the parameters $\theta$ and $\phi$. The Bloch sphere provides a very useful intuition when trying to understand the logic behind the one-qubit classification algorithms.
It should be noted that the surface of the Bloch sphere represents all \textit{pure} qubit states. A qubit might also be in a \textit{mixed} state, in which case, a third degree of freedom is added (corresponding to the length of the vector), and its state can be mapped to the interior of the Bloch sphere too.

A qubit's state is \textit{pure} when its state vector conveys a ``perfect'' knowledge of the qubit's state. For example, Equation \ref{eq:statevec} represents a pure state. However, there might be some uncertainty associated to the qubit's state, and it might not be possible to represent it with a unique state vector $|\alpha\rangle$. Instead, it will be a statistical combination of state vectors $ \{|\alpha\rangle$, $|\beta\rangle\}$ associated to probabilities $\{p_\alpha, p_\beta\}$. This is known as a \textit{mixed} state, and it can be represented with a \textit{density matrix}.
The density matrix of a mixed state consisting on an ensemble of states $|\psi_j\rangle$ with probabilities $p_j$ can be defined as:
\begin{equation}
    \rho \equiv \sum_j p_j |\psi_j\rangle \langle\psi_j|,
\end{equation}
where $\langle\psi_j|$ is the hermitian conjugate (also known as \textit{bra}) of $|\psi_j\rangle$ (which is referred to as \textit{ket}), so if $|\psi_j\rangle = \alpha_0 |0\rangle + \alpha_1 |1\rangle$, then $\langle\psi_j|= \alpha_0^* \langle0| + \alpha_1^* \langle1|$, where $\langle0| = \left( 1 \,\, 0 \right)$, $\langle1| = \left( 0 \,\, 1 \right)$, and $\alpha_i^*$ is the complex conjugate of $\alpha_i$. Note that a pure state is a sub-case of a mixed state where there is only one $|\psi_j\rangle$ and $p_j=1$.

Statistically, the density matrix can be seen as the co-variance matrix of the state distribution, and it is useful to determine the results of qubit measurements, as shown below.

\label{sec:gates}
Qubit states can be manipulated through unitary transformations in the form of \textit{quantum logic gates}. These transformations can be seen as $2\times2$ matrices, and correspond to different rotations in the Bloch sphere.
In its most general form, a single-qubit gate adheres to the following definition:
\begin{equation}
    U(\theta, \phi, \lambda) = \begin{pmatrix}
    \cos(\frac{\theta}{2}) & -e^{\mathrm{i}\lambda}\sin(\frac{\theta}{2})\\
    e^{\mathrm{i}\lambda}\sin(\frac{\theta}{2}) &
    e^{\mathrm{i}(\phi+\lambda)}\cos(\frac{\theta}{2})
    \end{pmatrix}
    \label{eq:u}
\end{equation}

Where different values of parameters $(\theta, \phi, \lambda)$ lead to an infinite number of possible gates, including some common ones such as the Pauli gates, namely $X=(\begin{smallmatrix}
    0 & 1\\
    1 & 0
    \end{smallmatrix})$, $Y=(\begin{smallmatrix}
    0 & -i\\
    i & 0
    \end{smallmatrix})$ and $Z=(\begin{smallmatrix}
    1 & 0\\
    0 & -1
    \end{smallmatrix})$, as well as the Hadamard gate $H=\frac{1}{\sqrt{2}}(\begin{smallmatrix}
    1 & 1\\
    1 & -1
    \end{smallmatrix})$.

These transformations are used in the context of quantum algorithms to perform the data encoding and processing steps, and they are characterized for being reversible.
A set of quantum gates is said to be \textit{universal} if any unitary transformation of the quantum data can be efficiently approximated arbitrarily well as a sequence of gates in the set.

At the final step of an algorithm, the qubits must be measured in order to read out the solution encoded in them. Quantum measurements suppose a crucial step when working with qubits and quantum states. They are the tool for retrieving the information stored within a quantum state, expressed in terms of probabilities. While a qubit contains an infinite amount of information in its state amplitudes, these amplitudes cannot be directly observed, and they have to be measured in order to retrieve this information. The amount of information that can be extracted depends on the probabilistic result of this measurement, but it is at most one bit per qubit, due to Holevo's bound \cite[Theorem 12.1]{nielsen2002quantum}.

A quantum measurement is mathematically described by a set of operators $\{M_a\}$ acting on the state space of the quantum system, and that satisfy $\sum_a M_a^\dagger M_a = \mathbb{1}$. The probability $P$ of a result (outcome) $a$ occurring when the state $|\psi\rangle$ is measured is:
\begin{equation}
    P(a) = \langle \psi | M_a^\dagger M_a | \psi \rangle.
\end{equation}

Once a measurement is performed, the state of the quantum system is immediately altered, and the new state after measurement is:
\begin{equation}
    | \psi' \rangle = \frac{M_a | \psi \rangle}{\sqrt{P(a)}},
\end{equation}
where $a$ is the concrete outcome obtained from the measurement.

One of the most common measurement strategies is the measurement on the computational basis, where the measurement operators correspond to the density matrices of the basis states, such that:
\begin{align}
    M_0 &= |0 \rangle \langle 0 |, \\
    M_1 &= |1 \rangle \langle 1 |.
\end{align}

Then, if we write the state of the qubit as $|\psi\rangle = \alpha_0 |0\rangle + \alpha_1 |1 \rangle$, the probabilities for each outcome correspond to:
\begin{align}
    P(0) = &\langle \psi | M_0^\dagger M_0 | \psi \rangle = \alpha_0^*\alpha_0 \langle 0 | 0 \rangle = |\alpha_0|^2,\\
     P(1) = &|\alpha_1|^2 = 1- |\alpha_0|^2.
\end{align}

Quantum algorithms can be implemented on quantum data through ordered sequences of quantum gates followed by measurements to retrieve information, composing \textit{quantum circuits}. A quantum circuit can be understood as a special sampling device that retrieves the result of a specific manipulation of a series of quantum states.
In a similar way to classical computing and machine learning, while quantum circuits were at the beginning designed as implementing a specific set of operations for executing a specific task \cite{shor1994,harrow2009quantum}, quantum circuits with trainable parameters that can be tuned for different tasks are becoming more and more popular (see, e.g., the review \cite{bharti2022}).

\section{The Single-Qubit Learning Algorithm}
\label{chap:algorithm}
According to Ref.~\cite{perez2021one}, \textit{a quantum neural network based on a single-qubit circuit can approximate any bounded complex function by storing its information in the degrees of freedom of a series of quantum gates}. This makes the single-qubit classifier a simple yet powerful tool for performing learning tasks.
This section will dive deeper into the construction of a machine learning algorithm that can be applied to classification problems based on the universal approximation properties of the single qubit, the principle of data re-uploading \cite{perez2020data}, the analogies that can be drawn to classical neural network architectures, and a review of the proof of universality from Ref.~\cite{perez2021one}.

\subsection{Data re-uploading and single-qubit QNNs}

The proposal of P\'{e}rez-Salinas et al.~\cite{perez2020data} is a QNN learning algorithm based on the concept of \textit{data re-uploading}, which allows to encode mathematical functions in the degrees of freedom of a series of gates applied to a single-qubit state.

QNN-based learning algorithms usually follow three distinct steps: Data loading (i.e., the encoding of the classical information into the state of a system of qubits), data processing, and measurement strategy (i.e., training).
Data re-uploading serves a double purpose within the single-qubit QNN: on one hand, the parametrized gates perform the encoding of classical data into the quantum space. On the other hand, these same gates allow to capture the underlying distribution of the data. In other words, data re-uploading allows to combine the steps of data loading and data processing in a single set of operations, minimizing the depth of the final circuit.

\subsubsection{Data re-uploading as encoding.}
In QNNs, there are several strategies employed for the data loading and processing steps. The most common alternative is to encode the input within the amplitudes of the statevector's wave function, for example, through \textit{basis encoding} ($U_\Phi: x \in \{0,1\}^n \rightarrow |x\rangle$) or \textit{amplitude encoding} ($U_\Phi: \vec{x} \in \mathbb{R}^N \rightarrow |\psi _{\vec{x}} \rangle= \sum^{N-1}_{i=0} x_{i}|i\rangle$),  but these strategies have been developed in the context of multi-qubit QNNs. When the quantum circuit is only composed of one qubit, these strategies become impractical, as a single qubit's statevector only contains two degrees of freedom to encode data.

Instead of relying on the amplitudes of the statevector to encode the data, the single-qubit QNN relies on a series of parametrized unitary rotations to capture the underlying properties of the data distribution. For example, in the case of a single-qubit classification task, the following feature map can be used instead to encode the data into the quantum feature space:
\begin{equation}
    U_\phi: \vec{x} \in \mathbb{R}^N \rightarrow |\psi\rangle = \mathcal{U}(\vec{\phi}, \vec{x})|0\rangle.
\end{equation}

\begin{figure}[ht]
    \centering
    \subfloat[\label{fig:nn}]{
        \includegraphics[height=0.3\textheight]{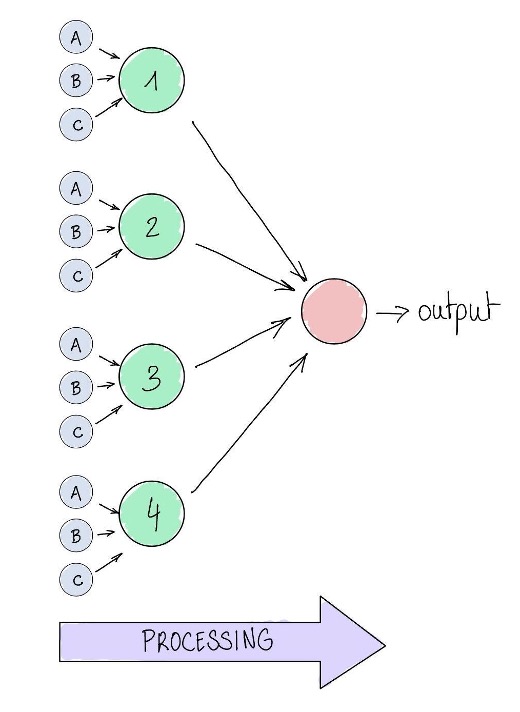}
    }
    \subfloat[\label{fig:qnn}]{
        \includegraphics[width=0.6\textwidth,clip,trim={0 0 0 -10cm}]{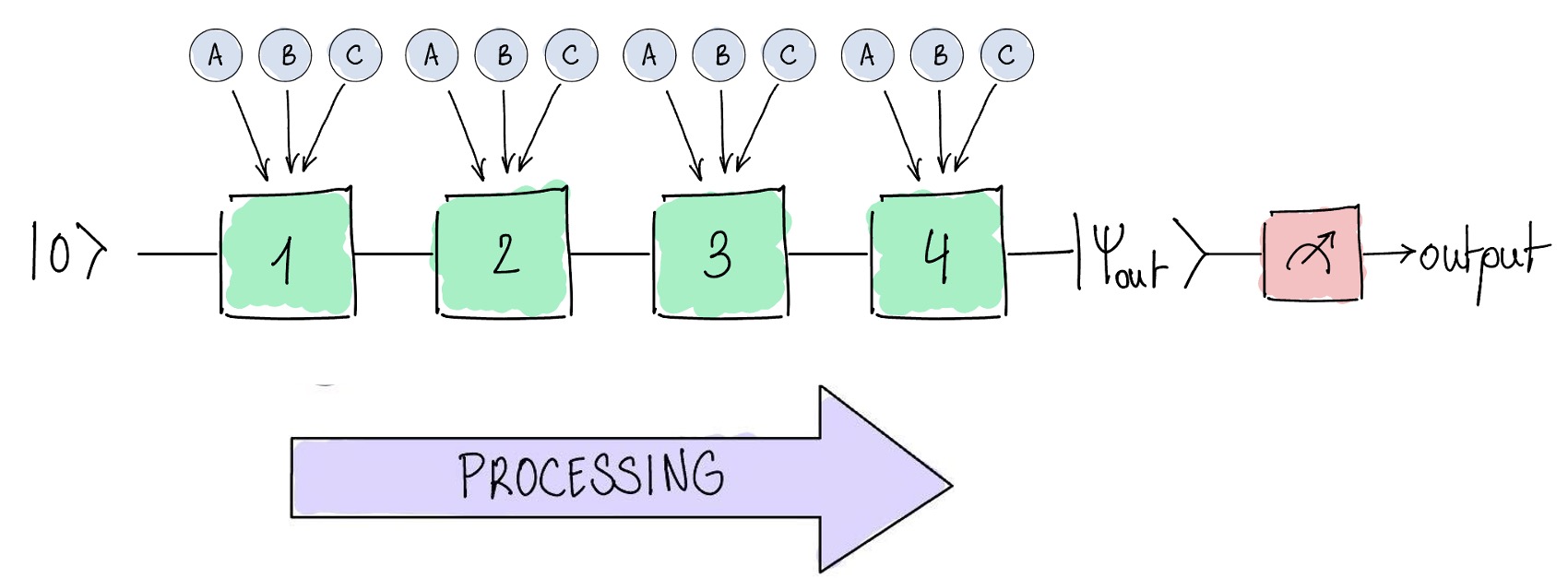}
    }
    \caption{Classical analogy for the single-qubit QNN architecture. \protect\subref{fig:nn} depicts a single-layer classical neural network (green) with an output neuron (pink). In this example, three inputs (labelled $A$, $B$, $C$) are fed to a single processing layer with four neurons (labelled from 1 to 4). Each neuron in the processing layer receives a copy of the inputs, processes it independently, and their outputs are combined in the output neuron. \protect\subref{fig:qnn} is the single-qubit QNN based on data re-uploading of Ref.~\cite{perez2020data}, where each layer (green) receives a copy of the inputs, and sequentially encodes the processing result in the state of a qubit that is then measured.}
    \label{fig:analogy}
\end{figure}

This strategy is inspired in the concept of classical neural networks, where the computations are performed by neurons organized in interconnected layers. In this quantum analogy, the unitary rotations $\mathcal{U}(\vec{\phi},\vec{x})$ can be seen to correspond to neurons, and form processing units that can be replicated to create layers. Each subsequent neuron ``re-uploads'' the classical input data, and is able to capture a particular feature of the distribution. Figure \ref{fig:analogy} shows the analogy between a single-hidden-layer classical NN and a single-qubit QNN.

\subsubsection{Data re-uploading as data processing structure.}
The use of data re-uploading situates single-qubit QNNs within the category of \textit{parametrized quantum circuits}, or \textit{variational quantum circuits}.
In variational quantum circuits, information is first encoded into a quantum state via an encoding strategy (also known in the machine learning literature as a \textit{feature map}), followed by a variational model (also known as \textit{ansatz}) that contain parametrized gates and performs the data processing step, and whose parameters are optimized for a specific task through loss function minimization after measurement.

In the case of the single-qubit QNN classification, the feature map and ansatz both are parametrized as arbitrary single-qubit rotations $U(\phi_1, \phi_2, \phi_3) \in SU(2)$. These rotations act as shown in Equation \eqref{eq:fm}, and the loss function can be computed after performing measurements on the resulting $|\psi\rangle$.
\begin{equation}
    \begin{aligned}
        \mathcal{U}(\vec{\phi}, \vec{x}) &\equiv U(\vec{\phi_N})U(\vec{x})\dots U(\vec{\phi_1})U(\vec{x})\\ |\psi\rangle &= \mathcal{U}(\vec{\phi}, \vec{x})|0\rangle
    \end{aligned}
    \label{eq:fm}
\end{equation}

In order to see a clearer correspondence to the classical analogy, we can introduce the concept of \textit{processing layer} as
\begin{equation}
    L(i) \equiv U(\vec{\phi_i})U(\vec{x}),
    \label{eq:ulayer}
\end{equation}
so that the final circuit corresponds to
\begin{equation}
\mathcal{U}(\vec{\phi}, \vec{x}) = L(N) \dots L(1).
\end{equation}

Given that each layer is composed of two unitaries, the total depth of a circuit with $N$ layers will be $2N$. According to the UAT (see Section \ref{sec:uat}), the more layers there are, the more representation capabilities will be present in the circuit. However, the more layers in a circuit the more time it will take to run. This may affect negatively the quality of the results due to the limited coherence times in current quantum processing units. For this reason, Ref.~\cite{perez2020data} also proposes a combined type of gate, where the input is fed to the circuit in a way closer to that of classical neural networks, namely:
\begin{equation}
    L(i)=U(\vec{\theta_i} + \vec{\omega_i} \circ \vec{x}),
    \label{eq:compressedulayer}
\end{equation}
where $\vec{\omega_i} \circ \vec{x} = (\omega_i^1x^1, \omega_i^2x^2,\omega_i^3x^3)$ is the Hadamard product of the input data vector and the vector of trainable weights. Figure \ref{fig:compressed_layer} shows graphically how this notation can reduce the total circuit depth.

However convenient this reduction may be, the gates used are able to handle a maximum of three dimensions. An extension to more dimensions can be achieved by adding extra gates (one gate per set of three dimensions), but this results in a new increase in computational complexity. In other words, with this formulation, the complexity of the circuit increases linearly with the size of the input space.

In order to circumvent this nuisance, Ref.~\cite{perez2021one} introduces a generalized version of the data re-uploading model: the fundamental UAT gate.

\subsubsection{Data re-uploading for a universal quantum approximant.}
As shown in Section \ref{sec:uat}, in classical machine learning the UAT states that a neural network with a single hidden layer can approximate any continuous function, and that the accuracy of the approximation increases the more neurons are added to this layer. In Ref.~\cite{perez2021one}, the authors propose a more generalized formulation for the data re-uploading layer that can be used to approximate any continuous function, the \textit{fundamental UAT gate}:
\begin{align}
    U^{UAT}(\vec{x}, \vec{\omega}, \alpha, \varphi) = R_y(2\varphi)R_z(2\vec{\omega}\cdot\vec{x} + 2\alpha),\quad
    \{\vec{\omega}, \alpha, \varphi\} \in \{\mathbb{R}^m, \mathbb{R}, \mathbb{R}\}.
    \label{eq:uatlayer}
\end{align}

\begin{figure}[ht]
    \centering
    \subfloat[\label{fig:original_layer} Original Unitary Layer]{
        \includegraphics[height=0.1\textheight]{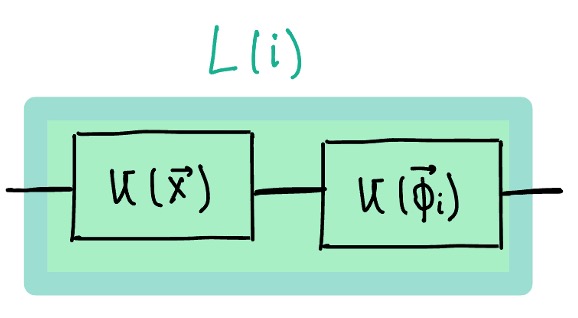}
    }
    \hfill
    \subfloat[\label{fig:compressed_layer} Compressed U. Layer]{
    \includegraphics[height=0.1\textheight]{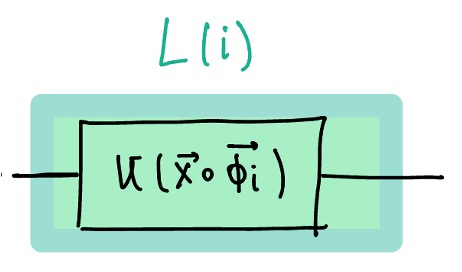}
    }
    \hfill
    \subfloat[\label{fig:uat_layer} UAT Layer]{
    \includegraphics[height=0.11\textheight]{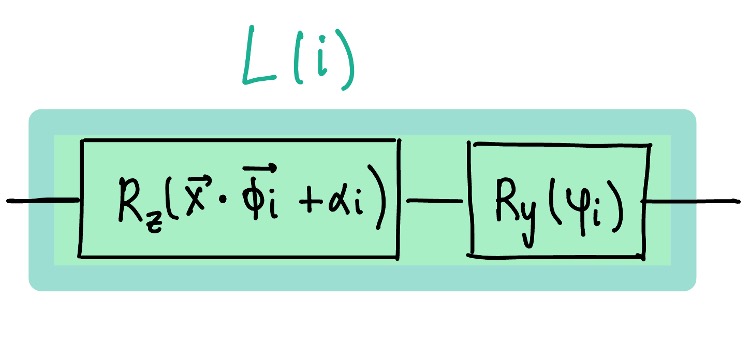}
    }
    \\[20pt]
    \subfloat[\label{fig:concat} Layer Concatenation Scheme]{
    \includegraphics[height=0.08\textheight]{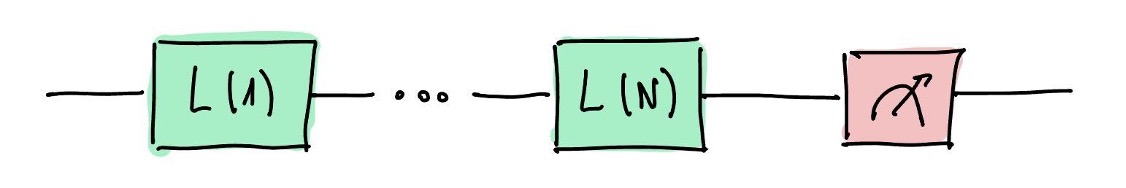}
    }
    \caption{Single-qubit QNN formulations from Ref.~\cite{perez2020data} and Ref.~\cite{perez2021one}. \protect\subref{fig:original_layer} is the original processing layer of Equation \eqref{eq:ulayer}, \protect\subref{fig:compressed_layer} is the compressed unitary layer of Equation \eqref{eq:compressedulayer}, and \protect\subref{fig:uat_layer} shows the UAT-based layer from Equation \eqref{eq:uatlayer}. These layers can be concatenated as illustrated in \protect\subref{fig:concat}.}
    \label{fig:layer_comparison}
\end{figure}

In the expression above, $R_z$ and $R_y$ represent rotation gates along the $Z$ and $Y$ axes, respectively, $\vec{\omega}$ is the vector of weights of the layer, $\alpha$ is the bias, and the role of $\varphi$ is that of the activation function (note that it enters in a rotation along the $Y$ axis, while the input data enters in a rotation along the $Z$ axis). This formulation allows to increase the size of the input space without any increase on the complexity of the circuit. Figure \ref{fig:layer_comparison} shows a side-by-side comparison of the three single-qubit layer formulations that have been introduced in this section.

\subsection{Training single-qubit QNNs}
\begin{figure}[ht]
    \centering
    \includegraphics[height=0.25\textheight]{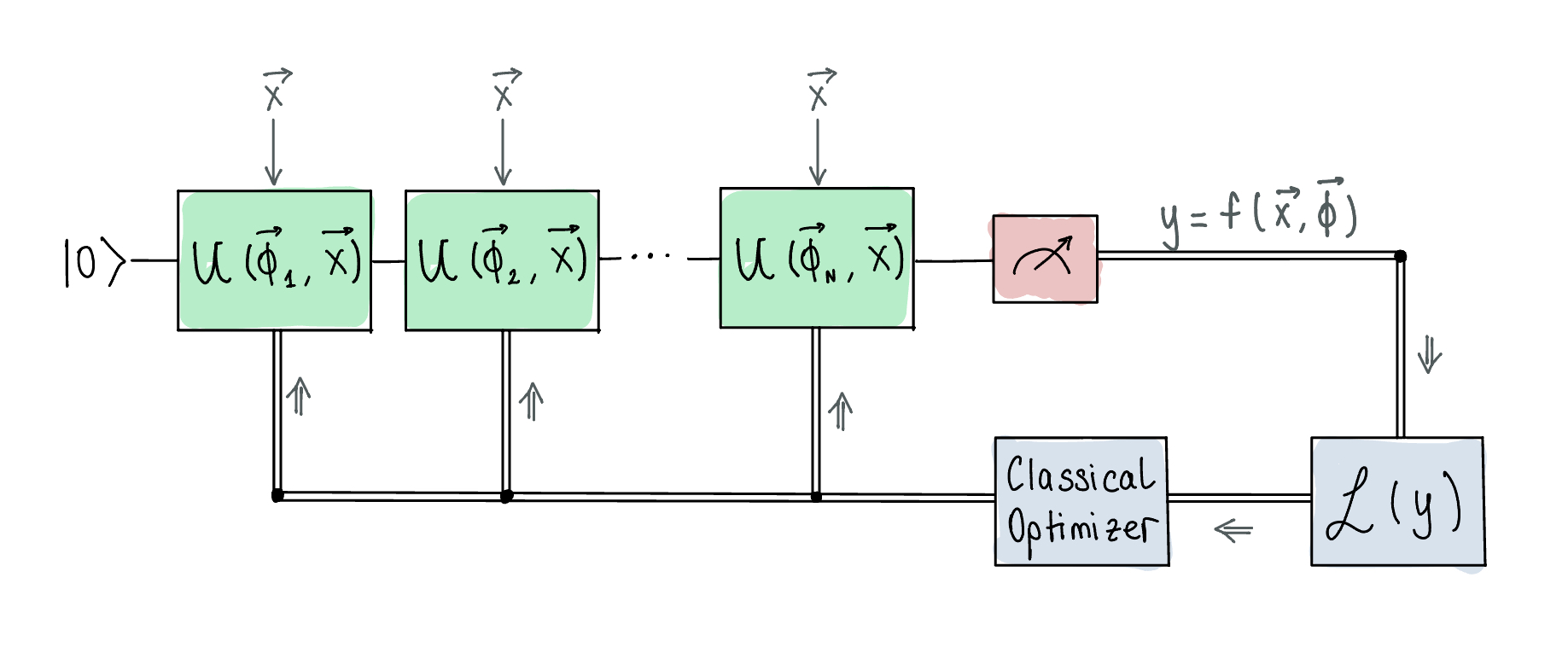}
    \caption{Training pipeline for a single-qubit QNN. Single lines denote the flow of the qubit, while double lines denote flow of classical information. Each of the blocks corresponds to one of the layers in Figure~\ref{fig:layer_comparison}. After the application of parameterized quantum gates to the qubit, this is measured and a cost function is computed classically. Gradients of this cost function are computed using a classical optimizer, and the parameters of the quantum gates are updated according to them.}
    \label{fig:training-qnn}
\end{figure}

Once the feature map and ansatz are defined for a target variational quantum circuit, it can be trained following the hybrid procedure illustrated in Figure \ref{fig:training-qnn}. The input data is loaded into the network with an initial set of arbitrary parameter values. The gates are applied and followed by a measurement operation at the end. The result of this measurement is fed into a specific cost function that is used to guide a classical optimizer to find the next set of parameters. This process is performed iteratively until the optimizer reaches the minimum cost.

The remainder of this section will focus on defining the missing algorithmic components (namely the measurement procedure, the cost function and the choice of classical optimizer) in order to perform a single-qubit QNN training for the specific case of a classification task, where the goal is to assign a discrete value (out of a small and finite number of possibilities) to each input, which represent its class.

\subsubsection{Measurement strategy.}
The goal of the quantum measurement strategy is to find the optimal way to associate the outputs from the quantum observations to the target classes. For these means, the training should facilitate the maximum distance between the different outputs. For example, in the case of binary classification, class A can correspond to state $|0\rangle$, and class B to $|1\rangle$. A way to measure this is to calculate the probability of obtaining the outcome $0$ when measuring the final state in the computational basis, and establish a threshold value $\lambda$ such that if $P(0)>\lambda$ the input data is associated to class A, and else for class B.

In terms of geometry, the classifier can be seen as a line dividing the Bloch sphere into two distinct sections. All states start pointing to the ``north pole'' (the state $|0\rangle$), and operations $L(i)$ correspond to rotations on the surface of this Bloch sphere, so that a set of parameters situates the state in the section corresponding to the right label. This means that an isolated data point could be correctly classified with a single unitary operation. However, a single rotation cannot be optimal for multiple distinct data points. Analogous to their classical counterparts, the addition of multiple layers in QNNs can solve this issue, allowing to define a more complex feature map without increasing the total number of qubits.

\subsubsection{Fidelity cost function.}
In the case of binary classification, the goal of the training is to force the output quantum states to be as close as possible to a particular state on the Bloch sphere ($|0\rangle$ or $|1\rangle$, depending on the label), maximizing the angular distance between two distinct targets. This angular distance can be measured through a metric known as the relative fidelity between two states. Given two density matrices, $\rho$ and $\sigma$, the fidelity is defined as the quantity:
\begin{equation}
     F(\rho ,\sigma ) =\left(\operatorname {tr} {\sqrt {{\sqrt {\rho }}\sigma {\sqrt {\rho }}}}\right)^{2},
\end{equation}
where $\text{tr}(z)$ is the trace operation, and it is calculated as the sum of elements of the main diagonal of a matrix.

When $\rho$ and $\sigma$ are pure quantum states, $\rho =|\psi _{\rho }\rangle \!\langle \psi _{\rho }|$  and $ \sigma =|\psi _{\sigma }\rangle \!\langle \psi _{\sigma }|$, the definition of fidelity is reduced to the overlap between states:
\begin{equation}
    F(\rho ,\sigma )=|\langle \psi _{\rho }|\psi _{\sigma }\rangle |^{2}.
\end{equation}

This allows to easily define a fidelity-based cost function:
\begin{equation}
    \chi_f^2(\vec{\theta}, \vec{\omega}) = \sum ^M_{\mu=1}(1-|\langle\psi_s^\mu|\psi(\vec{\theta},\vec{\omega}, \vec{x_\mu})\rangle|^2),
\end{equation}
where $\mu$ indicates the data point and $|\psi_s^\mu\rangle$ is the correct label state.

\section{Experiments and results}
\label{chap:experiment}
We have implemented single-qubit neural network experiments using \texttt{qiskit}, a Python-based \cite{python} open-source software development kit (SDK) developed by IBM Quantum \cite{Qiskit}. In particular, our implementation uses \texttt{qiskit-machine-learning}, a library for quantum computing and machine learning experiments. In addition to the elements in the library, we have prepared routines for the calculation of the fidelity loss function and defined classes for the three different single-qubit ansatz formulations that will be used in the experiments section.
The code for the experiments performed is available at \texttt{https://github.com/ElePT/single-qubit-qnn}.

We now summarize the experiments performed with the goal to provide a practical assessment of the power of a single-qubit classification algorithm for a real-world QML task, namely the detection of fraud in credit card transactions. All experiments were executed using \texttt{qiskit} on IBM's \texttt{aer\_statevector\_simulator}.

This section includes two subsections. In the first one, we use a toy dataset to analyze the properties of the different types of layers and choose the most effective for later use.
The second corresponds to the practical real-world scenario, where the applicability of the single-qubit quantum classification algorithm previously selected is tested using real-world data for a credit card fraud classification task.
\subsection{Model selection on the circle dataset}
\begin{figure}[ht]
    \centering
    \includegraphics[width=0.85\textwidth]{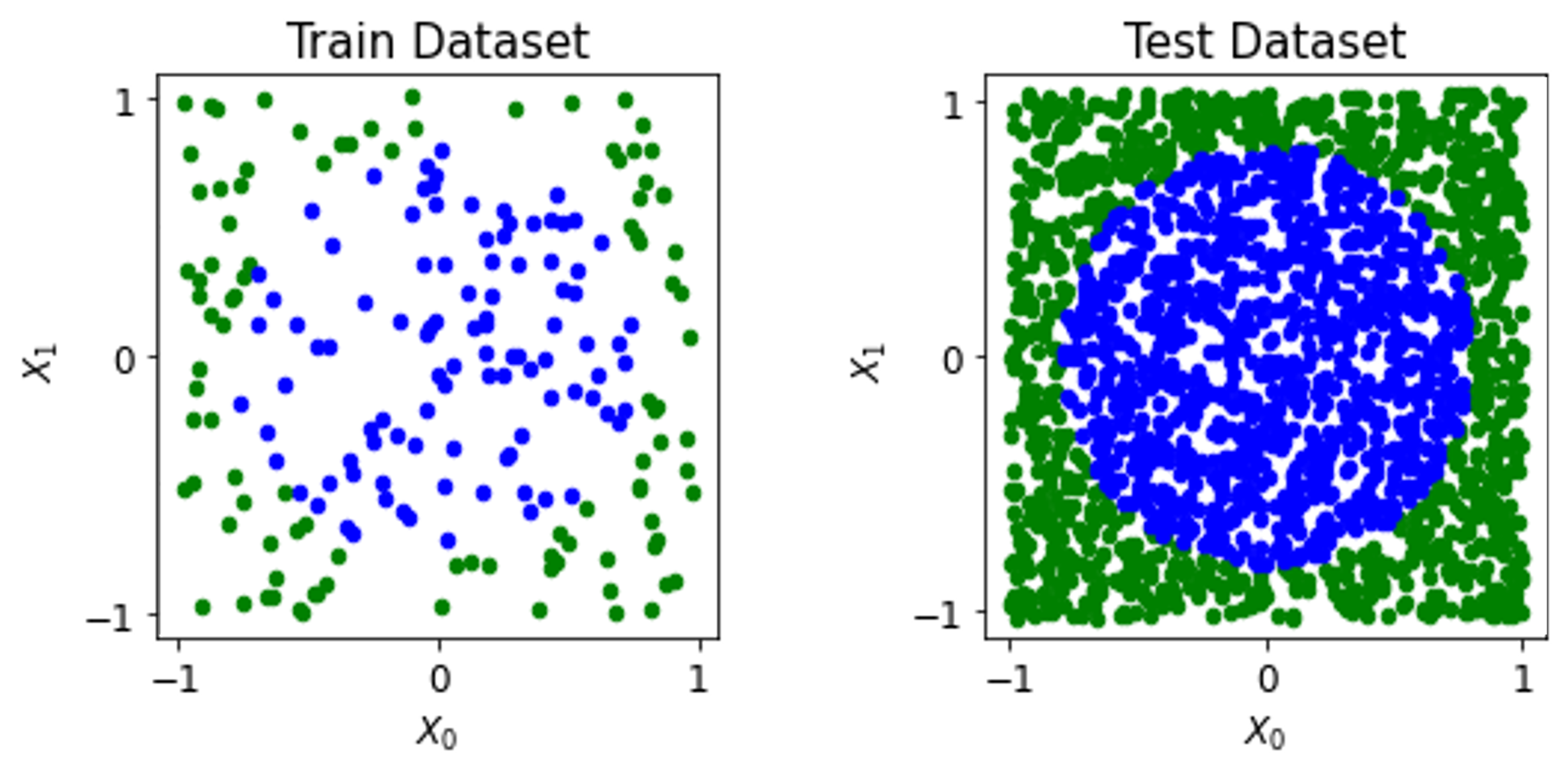}
    \caption{Synthetic datasets used in this section. They consist of uniformly distributed points $(x_0,x_1)\in[-1,1]^2$, and the label assigned corresponds to whether $x_0^2+x_1^2<1$ (blue) or not (green).}
    \label{fig:circle}
\end{figure}

Section \ref{chap:algorithm} presented three alternative ansatz formulations for the single-qubit classifier from \cite{perez2020data} and \cite{perez2021one}, namely the unitary-based layer of Equation \eqref{eq:ulayer}, the compressed unitary-based layer of Equation \eqref{eq:compressedulayer}, and the UAT-based layer of Equation \eqref{eq:uatlayer}. In the theoretical study of Ref.~\cite{perez2021one}, the UAT-based layer presents the advantage of providing a formulation where the number of gates does not increase linearly with the dimension of the data. In the scenario where all three formulations are assumed to be equally effective, it would make sense to favour this type of layer against the remaining alternatives. However, this might not necessarily be the case in a real-world experimental setup, where optimization dynamics might favour a specific neural network architecture.
While Refs.~\cite{perez2020data,perez2021one,perez2022ion} include numerical experiments, these experiments are not performed on the same dataset, and therefore provide little help when choosing the best layer architecture. For this reason, the following section will focus on addressing the question of optimal ansatz formulation choice, using a toy dataset introduced in Ref.~\cite{perez2020data}: the circle dataset depicted in Figure \ref{fig:circle}.

The circle dataset is a normalized synthetic 2-D classification dataset where points within a circle of a specific radius are labeled as one class (i.e, $1$), while the outside corresponds to the opposite class (i.e, $0$). The dataset is generated so that the labels are balanced. The training dataset is generated using 200 points, while the testing dataset contains 2000 points.

\subsubsection{Experiment 1: Comparison of layer formulations.}
The goal of this first experiment is to determine whether all three formulations are equally effective in a practical setting. For these means, the three ansatze introduced in previous sections are used to train QNNs on the circle dataset. According to the UAT (and as we will see in later experiments), the number of layers is a determining factor in the definition of the final accuracy. For this reason, the three models are trained with an equal number of layers (namely three). All trainings are performed with the LBFGS optimizer included in \texttt{qiskit} for the same number of epochs (50), on equal data instances, and the resulting accuracies are presented in Table \ref{tab:exp1}.

\begin{table}[ht]
    \centering
    \caption{Qualitative results for study of influence of layer formulation.}
    \begin{tabular}{||c | c c c | c c||}
     \hline
     \textbf{Layer Type} & \textbf{\#Layers} & \textbf{Depth} & \textbf{\#Params.}  & \textbf{Train Acc.} & \textbf{Test Acc.} \\ [0.5ex]
     \hline\hline
     Unitary & 3 & 6 & 9 &  \textbf{0.965} & \textbf{0.929} \\
     \hline
     Compressed Unitary & 3 & 3 & 18 & 0.965 & 0.909 \\
     \hline
     UAT & 3 & 6 & 15 & 0.94 & 0.88 \\
     \hline
    \end{tabular}
    \label{tab:exp1}
\end{table}

\begin{figure}[ht]
    \centering
    \includegraphics[width=\textwidth]{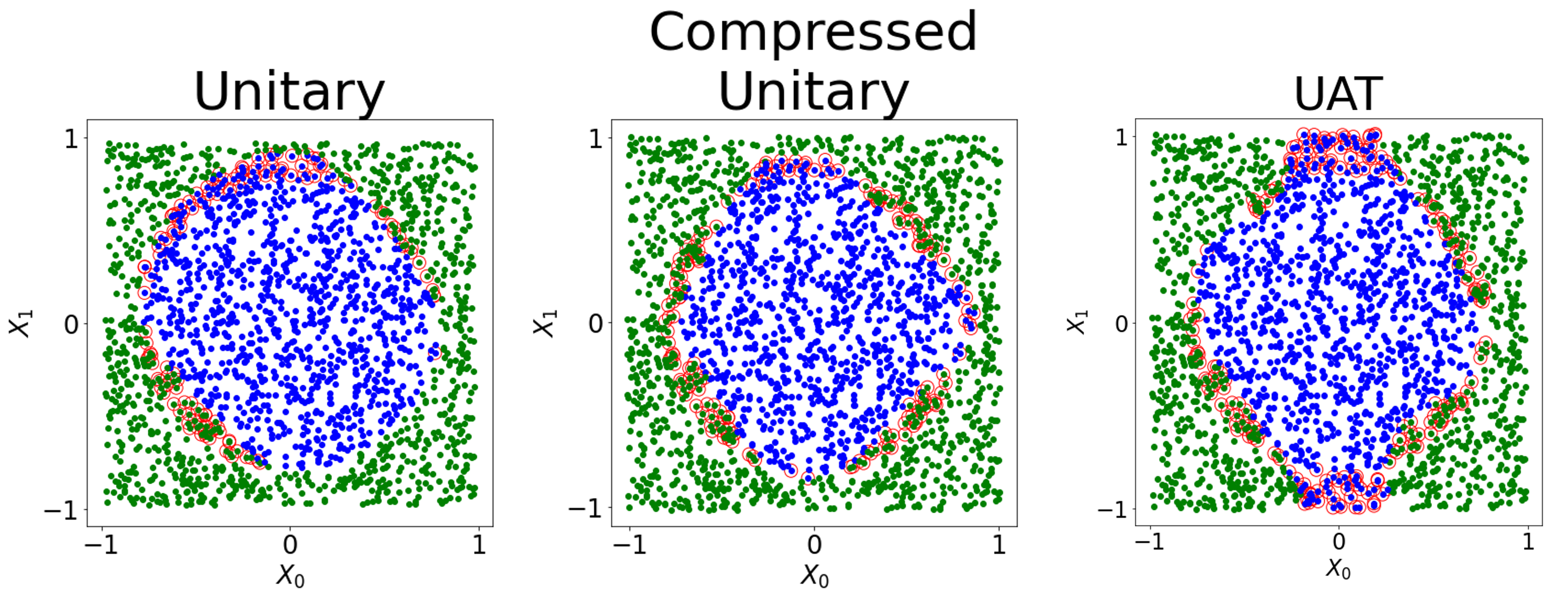}
    \caption{Analysis of the effect of the different types of layers in the classification accuracy. Points surrounded by circles represent misclassifications. Notably, the UAT layer seems not to correctly capture the shape of the dataset (figure shows the results on the test set).}
    \label{fig:exp1}
\end{figure}

While the difference between training accuracies is not very significant, the test accuracy appears to be up to 5\% lower for the UAT formulation. This seems to indicate that, for this number of layers, the generalization ability of QNNs based on UAT layers is worse than when using other layers. This can be seen explicitly in Figure \ref{fig:exp1}, where it is apparent that the UAT layer seems to fail at capturing the round shape of the boundary as accurately as the other two types of layers.

It has been noted in Ref.~\cite{perez2020data} that the unitary layer formulation is particularly well suited for symmetrical data distributions, such as the circle dataset. This opens the possibility for the UAT layer to perform better in more complex settings.

Because the UAT layer is composed exclusively of $R_z$ and $R_y$ rotations, despite it being theoretically universal, there is a chance that the optimizer might get stuck in a local minimum and fail to fully learn the underlying distribution of the data. The following experiment focuses on potential alternatives to overcome this drawback.

\subsubsection{Experiment 2: Adding data loading to the UAT layer.}
As explained in the previous section, the UAT layer is believed to be the most practically advantageous formulation in theory, due to its depth being independent of the number of data dimensions. However, in the circle dataset this does not seem to be the case. Part of the reason for this may be the fact that the first gate of the UAT layer is an $R_z$ gate, whose application on the state $|0\rangle$ is trivial (i.e., $R_z(\theta)|0\rangle=|0\rangle$ for any value of $\theta$). Since the initial states of the qubits in a QNN is precisely $|0\rangle$, the initial UAT layer in a QNN only contributes effectively with one parameter (the nonlinearity $\varphi$) to the network, and no information about the input data is used. In order to boost the training process with UAT, an initial data loading layer might be of help.

An initial hypothesis is that a data preparation step using a Hadamard gate would help take advantage of the state of superposition for facilitating the parameter search. Following this train of thought, it could be even more beneficial to perform the data preparation state with a generic parametrized unitary gate (recall Equation \eqref{eq:u}), whose parameters are optimized jointly with those of the ansatz.
Table \ref{tab:exp2} summarizes the result from this second experimental setup, where the UAT layer has been trained with different data loading routines. The results show a slight improvement in accuracy when using the generic unitary data loading step, while the Hadamard data loading does not seem to provide any advantage in this scenario.

\begin{table}[ht]
    \centering
    \caption{Accuracy study of influence of data loading step in UAT layers.}
    \begin{tabular}{|| c | c c c | c c||}
     \hline
    \textbf{Initial Preparation} & \textbf{\#Layers} & \textbf{Depth} & \textbf{\#Params.}  & \textbf{Train Acc.} & \textbf{Test Acc.} \\ [0.5ex]
     \hline\hline
     None & 3 & 6 & 15 & 0.94 & 0.88 \\
     \hline
     H & 3 & 7 & 15 & 0.93 & 0.88 \\
     \hline
     U & 3 & 7 & 18 & \textbf{0.95} & \textbf{0.89} \\
     \hline
    \end{tabular}
    \label{tab:exp2}
\end{table}

\begin{figure}[ht]
    \centering
    \includegraphics[width=\textwidth]{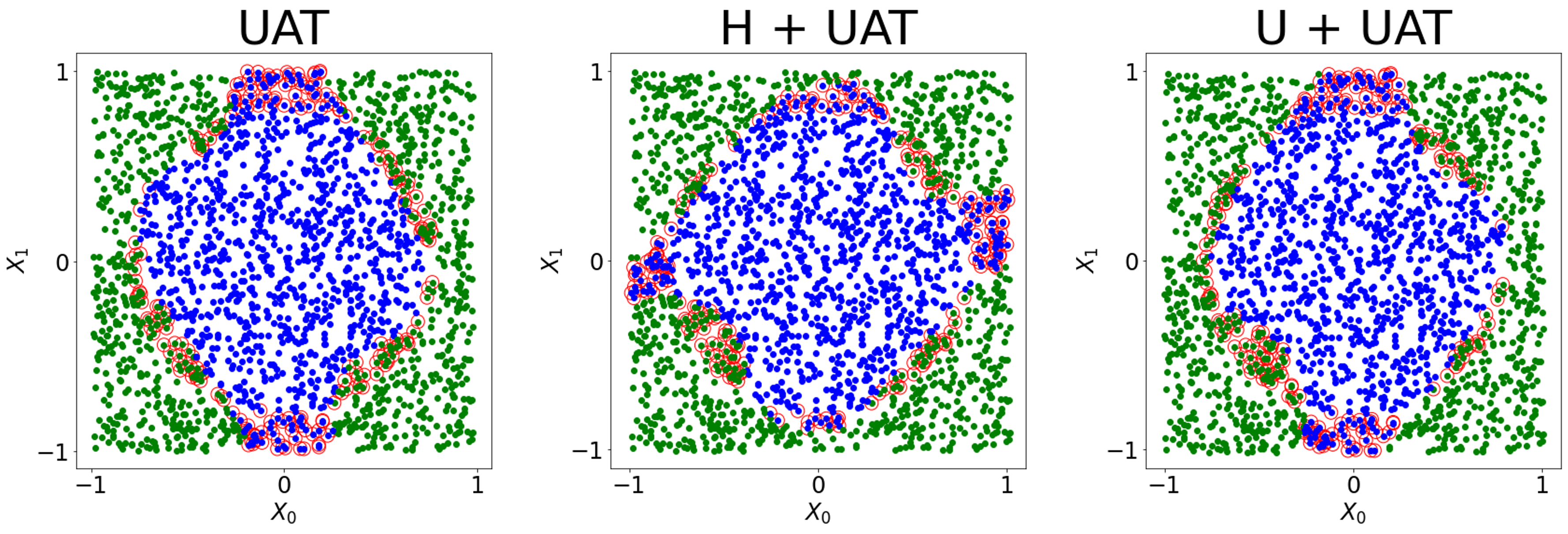}
    \caption{Qualitative results for study of influence of data loading step. Circled points represent misclassifications. In this case, none of the models appears to completely capture a round structure. This indicates a lack of expressive power that can be solved by adding more layers to the network.}
    \label{fig:exp2}
\end{figure}

The qualitative results from Figure \ref{fig:exp2} show that the unitary data loading does not seem to help a great deal in capturing the ``roundnesss'' of the circle. This might mean that this is not a data preparation issue, but rather a matter of the optimal number of layers necessary for the UAT to learn this particular data structure.

\subsubsection{Experiment 3: Selection of number of layers.}
\begin{figure}[ht]
    \centering
    \includegraphics[width=\textwidth]{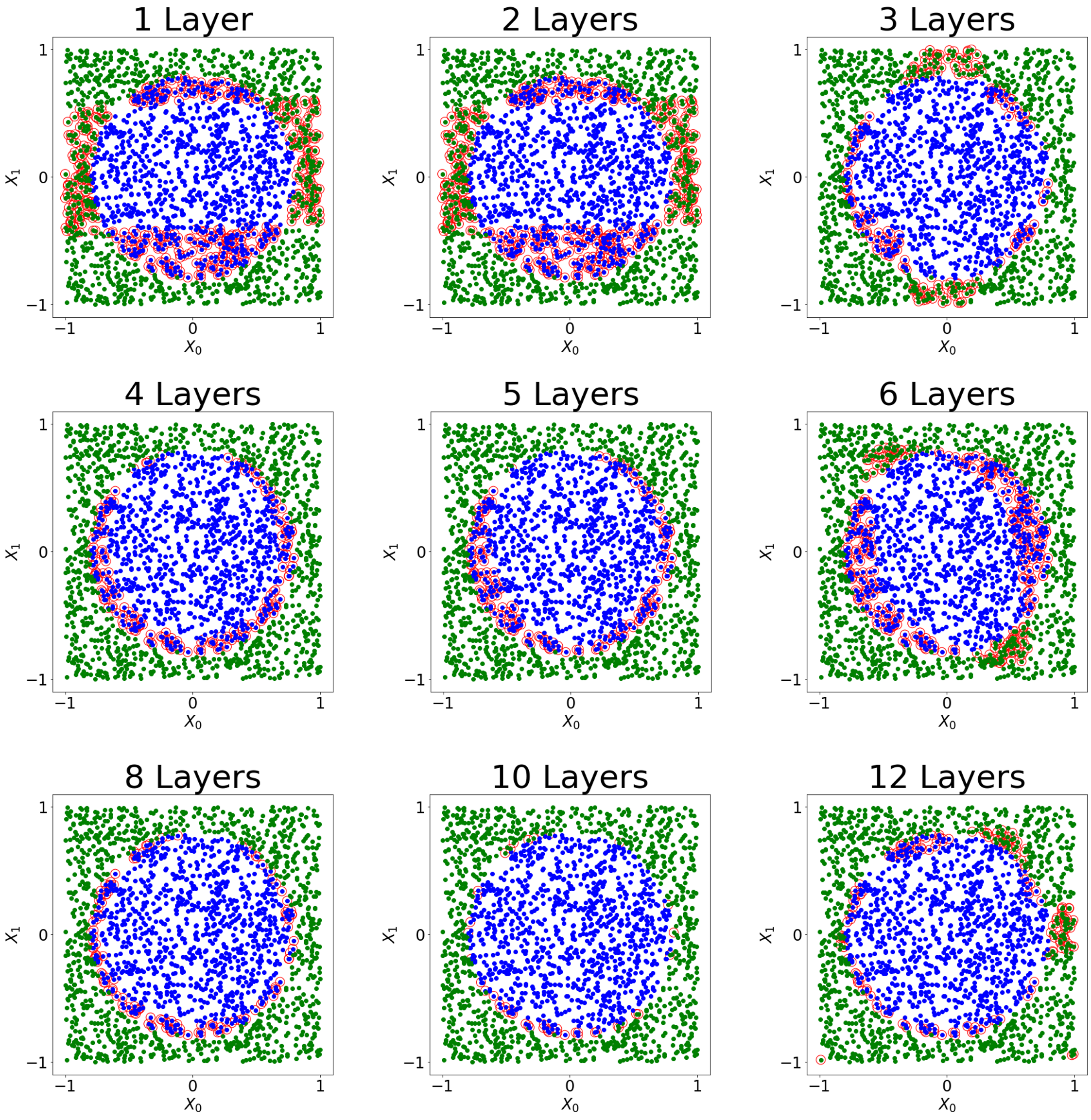}
    \caption{Qualitative results for study of influence of number of layers. Circled points represent misclassifications. The evolution towards learning a circular structure is explicit for low number of layers, achieving an optimal configuration for 4 layers.}
    \label{fig:exp3}
\end{figure}

This final experiment focuses on observing the effect of the number of layers in the efficacy of the UAT layer for the circle dataset. For these means, the same testing environment is reproduced for up to 12 layers. In all cases the remaining variables, namely the layer type (UAT) and the data loading step (unitary) are kept the same. The results are summarized in Table \ref{tab:exp3} and Figure \ref{fig:exp3}.

\begin{table}[ht]
    \centering
    \caption{Accuracy study of influence of number of layers in QNNs with UAT layers and unitary initial preparation.}
    \begin{tabular}{|| c c c | c c||}
     \hline
      \textbf{\#Layers} &  \textbf{Depth} &  \textbf{\#Params.}  &  \textbf{Train Acc.} &  \textbf{Test Acc.} \\ [0.5ex]
     \hline\hline
     1 & 3 & 8 & 0.74 & 0.74\\
     \hline
     2 & 5 & 13 & 0.74 & 0.73 \\
     \hline
     3 & 7 & 18 & 0.95 & 0.88 \\
     \hline
     \textbf{4} & \textbf{9} & \textbf{23} &  \textbf{0.93} &  \textbf{0.90} \\
     \hline
     5 & 11 & 28 & 0.94 & 0.90 \\
     \hline
     6 & 13 & 33 & 0.81 & 0.81 \\
     \hline
     8 & 17 & 43 & 0.92 & 0.93 \\
     \hline
     10 & 21 & 53 & 0.98 & 0.98 \\
     \hline
     12 & 25 & 63 & 0.95 & 0.90 \\
     \hline
    \end{tabular}
    \label{tab:exp3}
\end{table}

In this study, the qualitative results shown in Figure \ref{fig:exp3} are particularly illustrative. It becomes evident that three layers is the minimum number required for starting to see a circular decision boundary, and that for four layers the adjustment of the decision boundary to both the training and test datasets is already very good.
From five layers onward there is no obvious gain in the quality of the models when adding layers: there are some configurations for which overfitting is clear (see, e.g., the cases with six and twelve layers), while there are others, like that with ten layers, that outperform the four-layer case. Because the relation between increasing the number of parameters and the model's performance is not clear for large number of layers, and having more layers is an issue in current devices due to restricted coherence times, in the following we choose four layers as the overall optimal configuration.

\begin{figure}[ht]
    \centering
    \includegraphics[width=0.7\textwidth]{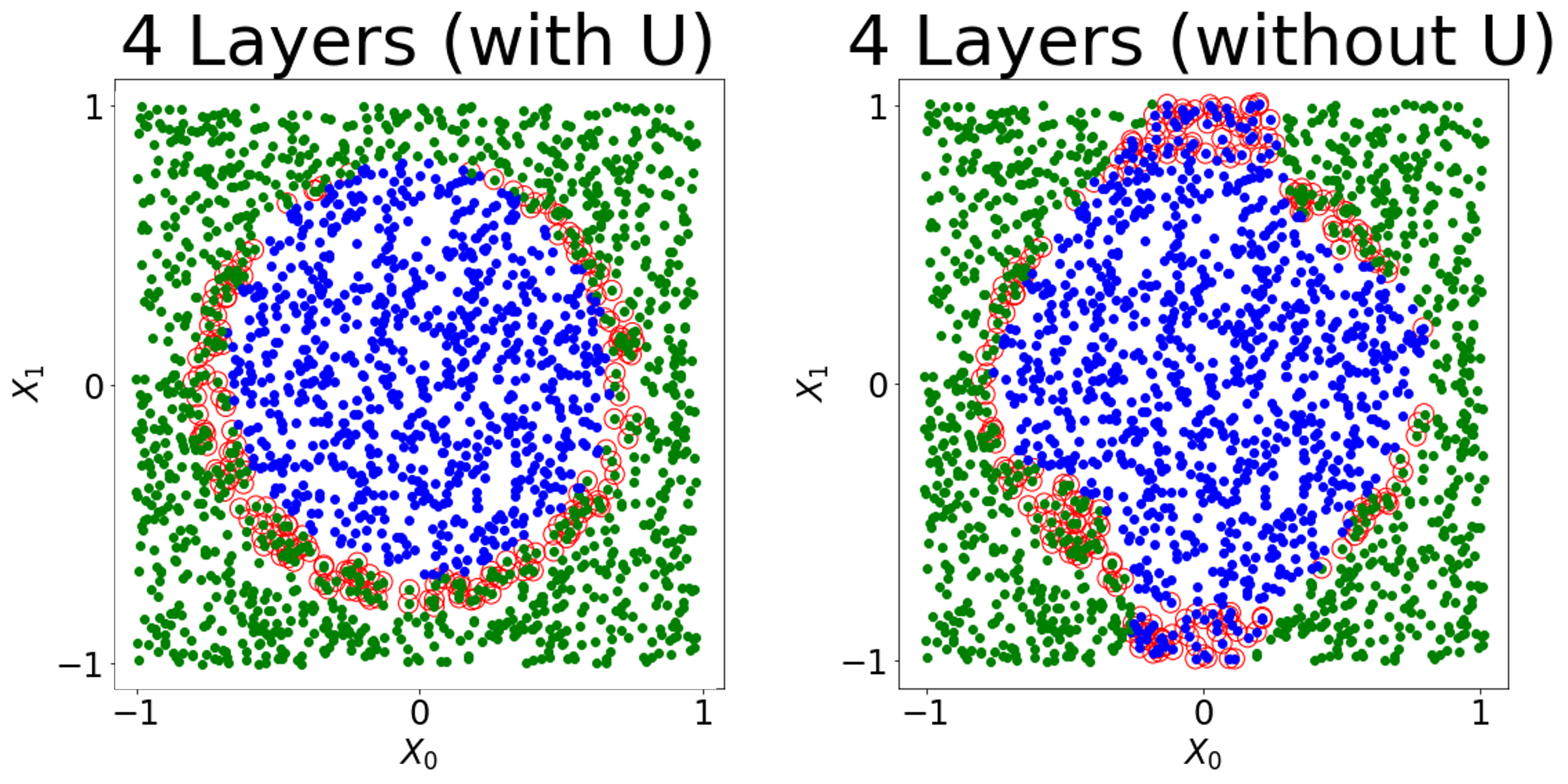}
    \caption{Comparison of optimal architectures for circle dataset. Circled points represent misclassifications. Adding an initial learnable rotation to the network clearly improves the ability to reproduce the features of the dataset.}
    \label{fig:exp3_bis}
\end{figure}

So far, the number of layers seems to be one of the most influential factors for determining the QNN's performance. Once the number of layers has been set to 4, it is possible to go back to experiment 2 and confirm the usefulness of the unitary data preparation step. Table \ref{tab:exp3_bis} and Figure \ref{fig:exp3_bis} provide a clear insight on this matter. When the number of layers is properly determined, an additional data processing step can boost training and help capture the inherent properties of the circle dataset.

\begin{table}[ht]
    \centering
    \caption{Confirmation of effectiveness of data loading step on UAT-based single-qubit QNNs, using the appropriate number of layers.}
    \begin{tabular}{|| c | c c c | c c||}
     \hline
      \textbf{Initial Preparation} &  \textbf{\#Layers} &  \textbf{Depth} &  \textbf{\#Params.}  &  \textbf{Train Acc.} &  \textbf{Test Acc.} \\ [0.5ex]
     \hline\hline
     \textbf{U} & \textbf{4} & \textbf{9} & \textbf{23} &  \textbf{0.93} &  \textbf{0.90} \\
     \hline
     None & 4 & 8 & 20 & 0.95 & 0.88 \\
     \hline
    \end{tabular}
    \label{tab:exp3_bis}
\end{table}

\subsection{Real-world dataset}
The focus of the previous section was to gain a better understanding of the practical behaviour of the single-qubit quantum classifier in a toy setting. The results of the experiments in that section suggest that the UAT formulation might not always lead to the best accuracies when used as-is, but that paired with a trainable data loading step it can lead to competitive results, while providing a circuit depth that does not increase with the input data dimension.

The goal of this section is to test out these hypotheses using a real-world, unstructured dataset. For these means, the data will be processed classically, and the single-qubit quantum classifier will be benchmarked against a classical neural network architecture.

\subsubsection{Credit card fraud detection dataset.}
The Kaggle credit card fraud detection dataset \cite{kaggle} contains transactions made by credit cards in September 2013 by European cardholders. There are a total of 284,807 transactions, out of which 492 are fraudulent. The dataset is highly unbalanced, the positive class (frauds) account for 0.172\% of all transactions, and the input data contains a total of 30 dimensions.
Recent works \cite{kyri2022frauddet} have analyzed fraud detection in the context of QNNs. However, the authors do not consider the case of single-qubit architectures, which is what we focus on in this section.

This volume of data is not compatible with the current state of NISQ hardware. For this reason, an initial step of classical pre-processing has been performed to adapt the data to the capacity of the classifiers while maintaining the essence of the problem.
The pre-processing has been executed in two steps:

\begin{enumerate}
    \item \textbf{PCA dimensionality reduction}: Employing the principal component analysis (PCA) technique for compressing high-dimensional data into a lower-dimensional feature subspace with the goal of maintaining most of the relevant information.
    \item \textbf{Data sampling}: Randomly selecting a subset of data points to reduce the total dataset size.
\end{enumerate}

\begin{figure}[ht]
    \centering
    \includegraphics[width=\textwidth]{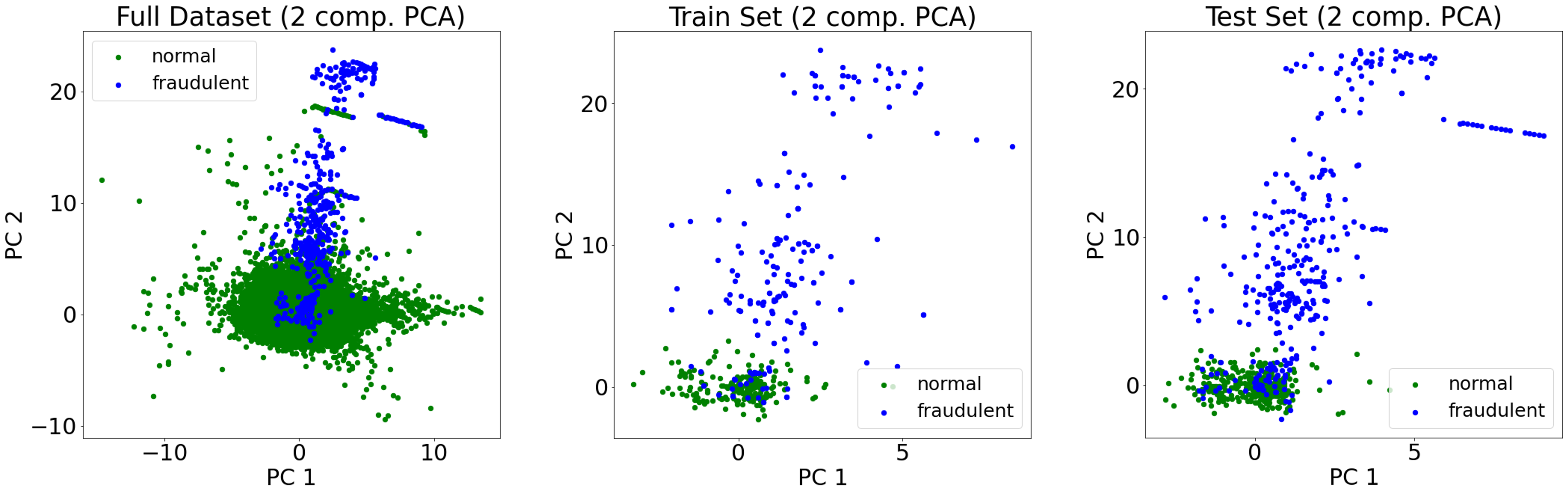}
    \caption{Credit card fraud detection dataset. The dataset is obtained by doing PCA to the Kaggle Credit Card Fraud dataset. Images show the (left) full, (center) training, and (right) test datasets. Green points correspond to non-fraudulent transactions, while bule points correspond to fraudulent ones.}
    \label{fig:pca}
\end{figure}

Step (i) allows to reduce the total number of workable features to two, as shown in Figure \ref{fig:pca}. Step (ii) serves a double purpose: on one hand, it helps to reduce the size of the training and test datasets to a manageable number of samples (400 each); on the other hand, it allows to tackle the initial class imbalance by forcing the class ratio to be close to 50-50. Imbalanced classification problems are challenging even in the context of classical machine learning, and dealing with this additional challenge has been deemed out of scope for the sake of these experiments. The result are the training and test sets shown in Figure \ref{fig:pca}.

\subsubsection{Establishing a classical baseline.}
In order to put in context the result of the quantum classification algorithms, we train classical counterparts to contrast against. These are two simple neural networks that have been written and trained using the Keras machine learning library in Python \cite{keras}.

\begin{figure}[ht]
    \centering
    \includegraphics[width=\textwidth]{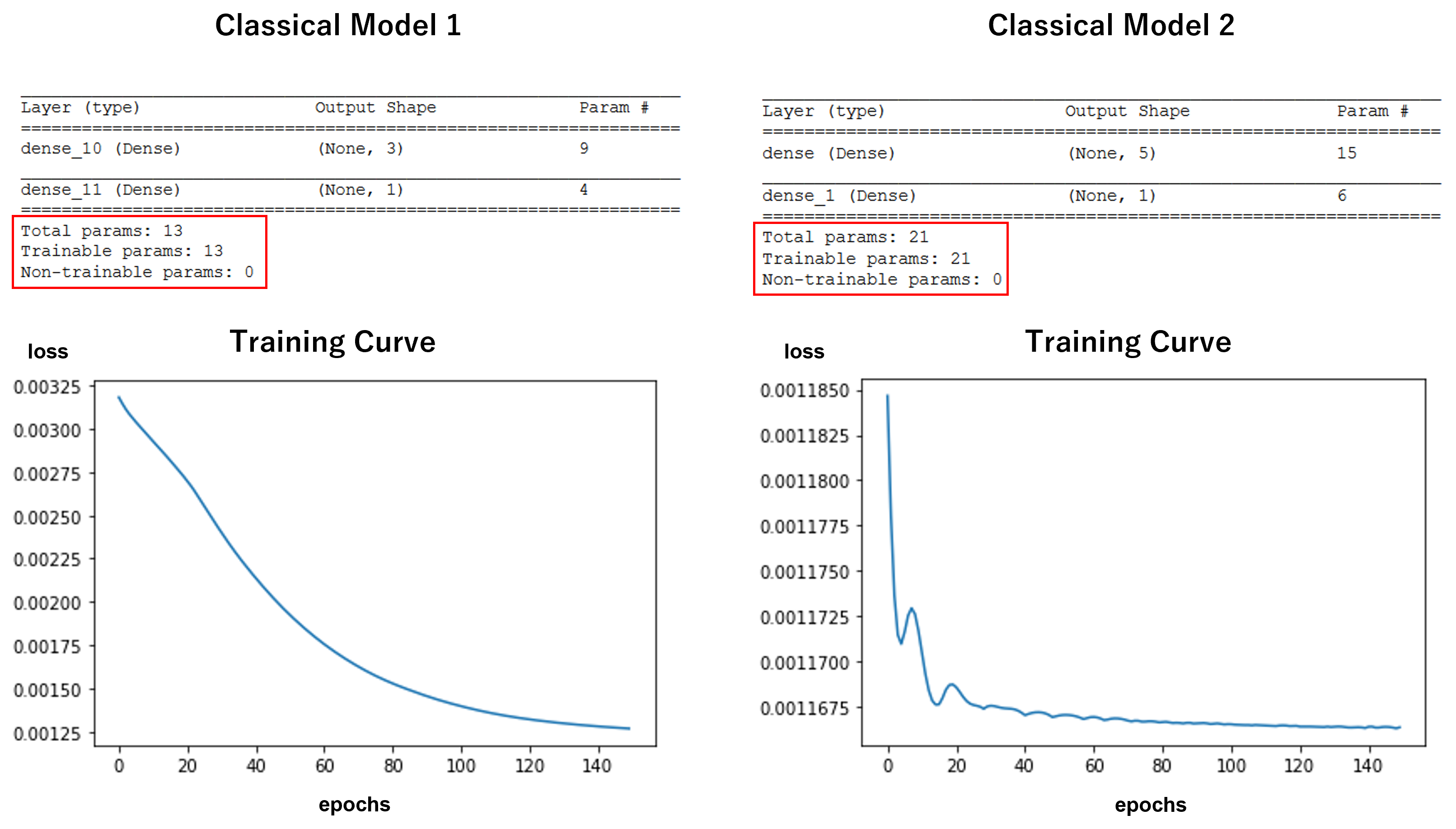}
    \caption{Results of the training of analogous classical models. We choose networks with a number of parameters similar to the QNNs (see Table 6), and train using standard techniques.}
    \label{fig:classical}
\end{figure}

The criteria for selecting the network size has been the total number of trainable parameters used. Both models have been trained using an Adam optimizer and L2 loss function for a total of 150 epochs. Figure \ref{fig:classical} shows the model description and training curves for the classical models. The results from the classical model trainings have been incorporated into Table \ref{tab:exp5}.

\subsubsection{Tests using the single-qubit QNN.}
Once the classical baseline has been set, it is the time for selecting the optimal single-qubit architecture. The set of experiments on the circle dataset suggest the following intuitions:

\begin{enumerate}
    \item The UAT formulation + U gate initialization has potential to be the best at generalizing.
    \item Tuning the number of layers is an important step of the process.
    \item In certain dataset distributions, the unitary layer formulation can surpass the UAT one.
\end{enumerate}

\begin{figure}[ht]
    \centering
    \includegraphics[width=\textwidth]{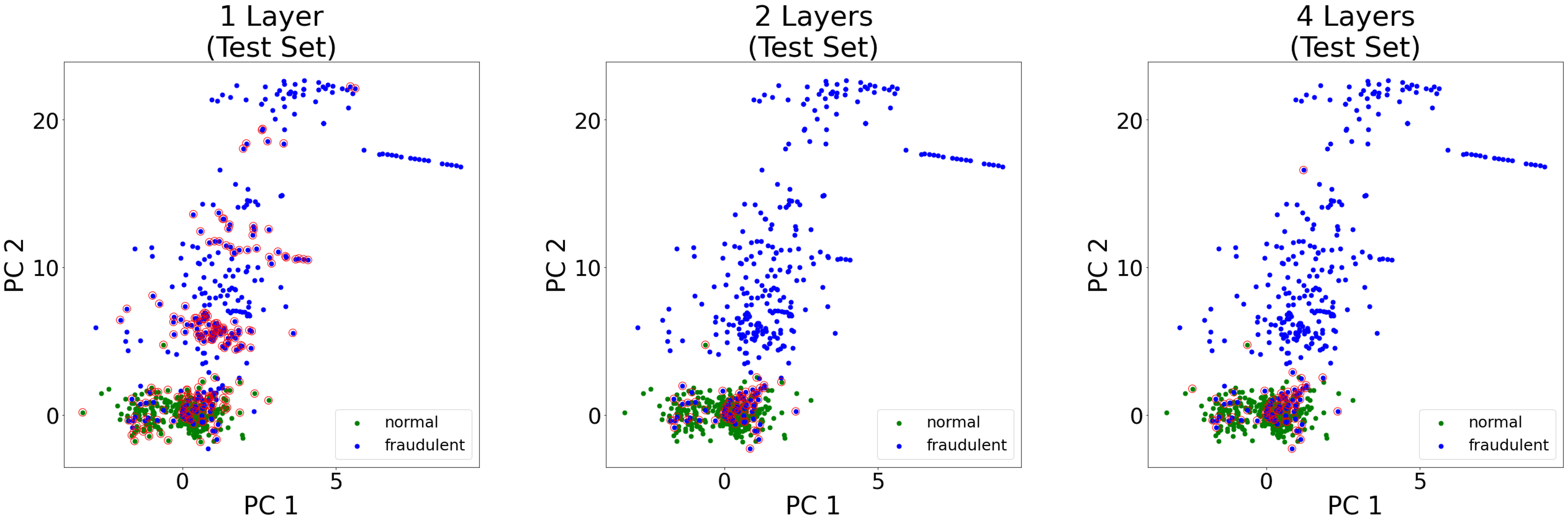}
    \caption{Qualitative study of QNN performance for different number of layers. Circled points represent misclassifications. While using just one layer produces a model with a periodicity that is not present in the training data, the additional parameters in subsequent layers correct this to provide more accurate models.}
    \label{fig:layers}
\end{figure}

With these guidelines in mind, the first set of experiments aims at finding the optimal number of layers for the (UAT + U) formulation. Figure \ref{fig:layers} shows that there is a qualitative leap in performance between 1 and 2 layers, but this difference becomes less noticeable when the number of layers continues growing. Table \ref{tab:layers} helps confirm this difference in a quantitative manner, by introducing the concepts of precision $P$ and recall $R$, where \textbf{tp}=\{true positives\}, \textbf{tn}=\{true negatives\}, \textbf{fp}=\{false positives\}, \textbf{fn}=\{false negatives\}, and:
\begin{align}
    A &= \frac{tp+tn}{tp+tn+fp+fn}, \\
    P &= \frac{tp}{tp+fp}, \\
    R &= \frac{tp}{tp+fn}.
\end{align}

\begin{table}[ht]
    \centering
    \addtolength{\leftskip} {-2cm}
    \addtolength{\rightskip}{-2cm}
    \caption{Results on performance metrics as a function of the number of layers on test data for the UAT + U formulation.}
    \begin{tabular}{||c| c c c c| c c c||}
     \hline
     \textbf{\#Layers} & \textbf{tp} & \textbf{tn} & \textbf{fp} & \textbf{fn} & \textbf{Acc.} & \textbf{Precision} & \textbf{Recall} \\ [0.5ex]
     \hline\hline
     1 & 187 & 276 & 147 & 54 & 0.697 & 0.56 & 0.78 \\
     \hline
     \textbf{2} & \textbf{269} & \textbf{327} & \textbf{65} & \textbf{3} & \textbf{0.898} & \textbf{0.81} & \textbf{0.99} \\
     \hline
     4 & 266 & 328 & 68 & 2 & 0.895 & 0.796 & 0.99 \\
     \hline
     8 & 264 & 328 & 66 & 2 & 0.897 & 0.80 & 0.99 \\
     \hline
     10 & 226 & 310 & 104 & 20 & 0.812 & 0.68 & 0.92 \\
     \hline
    \end{tabular}
    \label{tab:layers}
\end{table}

The results of Table \ref{tab:layers} suggest that the optimal number of layers for this problem is 2. Once this is fixed, we benchmark this optimal QNN against the following networks:
\begin{enumerate}
    \item Two classical NNs with an equivalent number of parameters
    \item Three control QNNs to confirm the assumptions from the previous section: one with unitary formulation, and two with UAT formulation and no data loading step.
\end{enumerate}

\begin{figure}[ht]
    \centering
    \includegraphics[width=\textwidth]{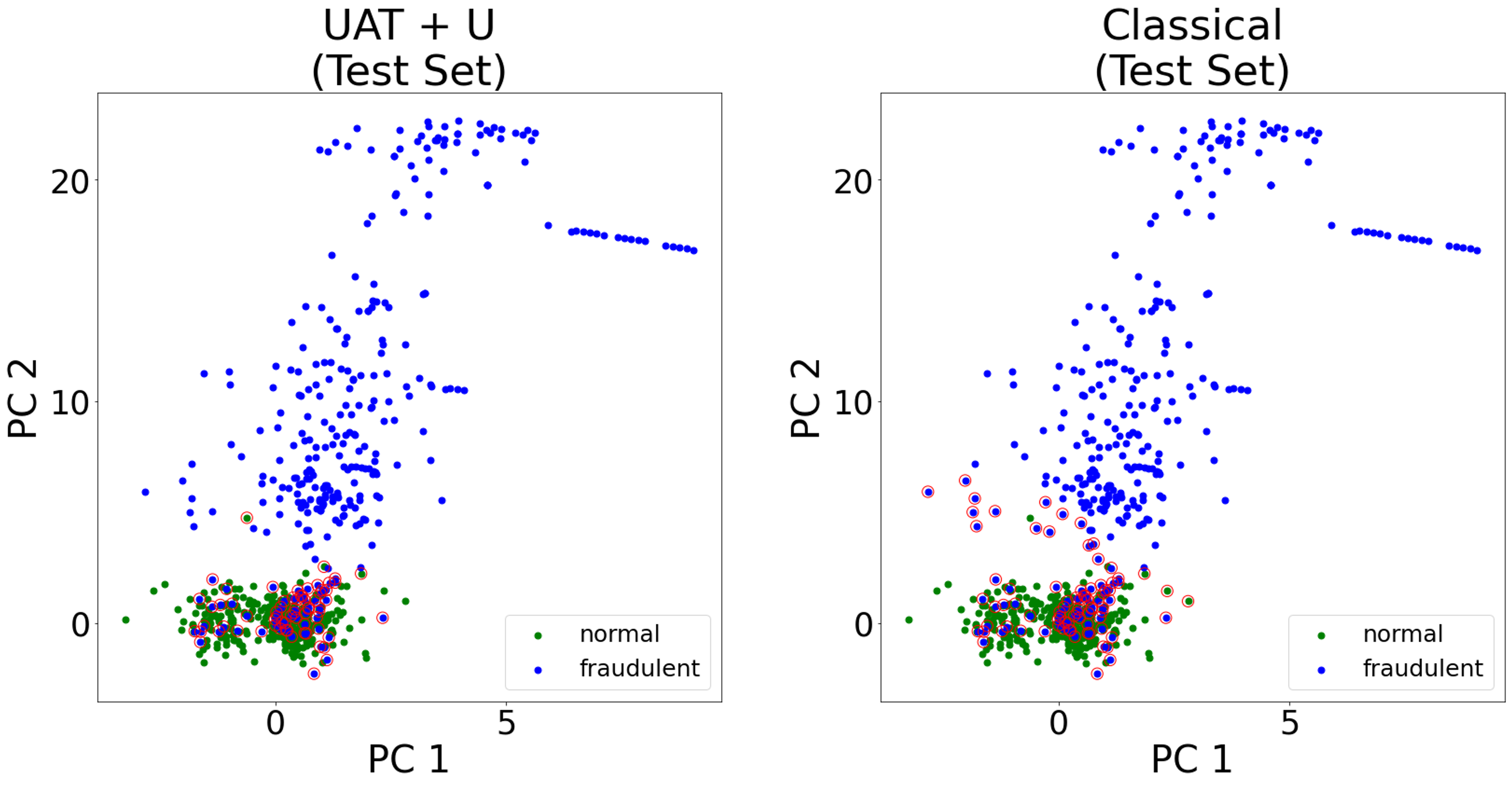}
    \caption{Qualitative comparison of best-performing QNN and classical counterpart. Circled points represent misclassifications. Notably, both models fail to classify almost the same points.}
    \label{fig:last}
\end{figure}

We have summarized the results from this benchmarking in Table \ref{tab:exp5}, which shows that the UAT formulation can indeed be comparable to a classical neural network of similar size, and by adding a data loading step, the performance of the network can increase. Figure \ref{fig:last} shows a side-by-side comparison of the test results for the best quantum formulation (highlighted in bold font in the table) and the corresponding classical neural network.

\begin{table}[ht]
    \centering
    \addtolength{\leftskip} {-2cm}
    \addtolength{\rightskip}{-2cm}
    \caption{Compilation of benchmarking results}
    \begin{tabular}{||c c | c c  | c c c||}
     \hline
     \textbf{L.Type} & \textbf{Initial Prep.} & \textbf{\#Layers} & \textbf{\#Params.} & \textbf{Acc.} & \textbf{Precision} & \textbf{Recall} \\ [0.5ex]
     \hline\hline
     UAT & U & 1 & 8 & 0.695 & 0.570 & 0.760\\
     \hline
     \textbf{UAT} & \textbf{U} & \textbf{2} & \textbf{13} & \textbf{0.902} & \textbf{0.810} & \textbf{0.990} \\
     \hline
     UAT & U & 4 & 23 & 0.897 & 0.803 & 0.990\\
     \hline
     UAT & U & 8 & 43 & 0.897 & 0.800 & 0.992 \\
     \hline
     UAT & U & 10 & 53 & 0.812 & 0.685 & 0.918 \\
     \hline
     \hline
     UAT & None & 2 & 10 & 0.830 & 0.688 & 0.961\\
     \hline
     UAT & None & 3 & 15 & 0.857 & 0.711 & 0.972\\
     \hline
     \hline
     Unitary & None & 3 & 9 & 0.688 & 0.624 & 0.715\\
     \hline
     \hline
     \textbf{Classical 1} & \textbf{None} & \textbf{2} & \textbf{13} & \textbf{0.898} & \textbf{0.797} & \textbf{1.00} \\
     \hline
     Classical 2& None & 2 & 21 & 0.895 & 0.806 & 0.982 \\
     \hline
    \end{tabular}
    \label{tab:exp5}
\end{table}

\section{Discussion and future work}
\label{chap:conclusion}
This paper has explored the applicability of a simple quantum neural architecture, composed of a single qubit and using data re-uploading techniques, to real-world data in the context of NISQ applications.

Instead of trying to prove a specific advantage, this work looks for the architecture that uses the minimal amount of quantum resources possible and is still able to solve a real-world task with results that are on-par with similar classical architectures. This architecture is the single-qubit universal quantum approximator proposed in Refs.~\cite{perez2020data} and \cite{perez2021one}, which has been discussed in detail in Section \ref{chap:algorithm}. The authors of these works propose a series of alternative layer formulations, but all of them can be trained in a variational manner by using a fidelity-based loss function.

In Section \ref{chap:experiment} we proposed a novel implementation of the single-qubit neural network using the \texttt{qiskit} SDK. We used this implementation in different variations for classification purposes in two different datasets, a toy classification problem with two-dimensional data where the class is determined by whether the datapoint lies inside the circle of radius 1 centered at the origin, and a real-world dataset of credit card transactions where the class is determined by whether the transaction is fraudulent or not.

Experiments on the toy dataset allow to gain further insights into the process of building a single-qubit quantum architecture. In particular, it shows the different accuracies associated to different layer formulations, the use of an initial data loading layer and different number of layers for a specific problem. These can all be considered hyperparameters to be optimized when designing a quantum neural network training scheme.

The insights gathered on the initial experiments help on the second part of the experimental procedure, where a single-qubit classification architecture is benchmarked against a classical counterpart on a credit card fraud detection task. The results from this second set of experiments show that single-qubit classifiers can achieve a performance on-par with classical counterparts under the same set of training conditions.

These results are promising, but should not be taken as a proof of advantage. In this type of scenarios, classical machine learning still holds the ``upper hand'', as it can deal with large amounts of data in the context of deep neural networks. However, if one qubit is indeed able to learn the necessary relations between data and associated labels in an effective manner, future architectures involving multiple qubits could hone specific quantum properties, such as entanglement, that could potentially lead to an advantage.

More specific future works could include the extension of these practical architectures beyond binary classification problems. As, in theory, single qubits should be able to hand multi-class classification problems with different measurement strategies, discussed in Ref.~\cite{perez2020data}.

\section*{Acknowledgments}
This work is supported by the Spanish Ministry of Science and Innovation MCIN/AEI/10.13039/501100011033 (CEX2019-000904-S and PID2020-113523GB-I00, also funded by ERDF ``A way of making Europe''), the Spanish Ministry of Economic Affairs and Digital Transformation (project QUANTUM ENIA, as part of the Recovery, Transformation and Resilience Plan, funded by EU program NextGenerationEU), Comunidad de Madrid (QUITEMAD-CM P2018/TCS-4342), and the CSIC Quantum Technologies Platform PTI-001.

\section*{References}

\bibliographystyle{unsrt}
\bibliography{mybib}

\end{document}